# Simulating e-Commerce Client-Server Interaction for Capacity Planning

## Ilija S. Hristoski*, Pece J. Mitrevski**

\* "St. Clement of Ohrid" University, Faculty of Economics, Prilep, Macedonia
\*\* "St. Clement of Ohrid" University, Faculty of Technical Sciences, Bitola, Macedonia

**Abstract**—Contemporary ways of doing business are heavily dependent on the e-Commerce/ e-Business paradigm. The highest priority of an e-Commerce Web site's management is to assure pertinent Quality-of-Service (QoS) levels of their Web services continually, in order to keep the potential e-Customers satisfied. Otherwise, it faces an immense possibility of losing both e-Customers and revenues, along with a big possibility of gaining bad reputation due to either poor performance or unavailability of the e-Commerce Web site. In order to avoid numerous unpleasant consequences, by designing and implementing e-Commerce Web sites that will always meet e-Customer's high expectations, a relevant performance models have to be utilized to obtain appropriate performance metrics. A continuous assessment of current performances and especially predicting future needs are the subjects of capacity planning methodologies. Within this paper, such a predictive model has been described and evaluated by using discrete-event simulation of both the client-side and server-side processes involved. In addition, the paper discusses the performance metrics obtained as a function of the intensity and quality of the workload parameters.

## I. INTRODUCTION

Keeping e-Customers of a particular e-Commerce Web site satisfied has always been a task of a highest priority for its management. E-Customers' dissatisfaction can lead to company's bad reputation, loses of both current and potential e-Customers, and substantial financial loses. Besides other leading factors for e-Customers' dissatisfaction, poor Web site performance is ranked second, just behind high product prices and shipping costs [1]. Since most Internet users have broadband connections today, poor Web site performance, including even Web site crashes, have been considered the main generators of e-Customers' dissatisfaction. According to Hoxmeier & DiCesare, user satisfaction is inversely related to response time, which 'could be the single most important variable when it comes to user satisfaction' [2]. In the virtual world of Internet, where time is money, speed is the only one-dimensional criterion that matters, whilst the expression 'faster is better' is the simplest equation in the entire field of Internet strategy [3]. Regarding the response time, the famous '8-seconds rule' has already changed. Four seconds is now considered the maximum length of time the average online shopper will wait for a Web page to load in a browser, before potentially abandoning the retail site, with a tendency to be halved again in the forthcoming years [1]. In order to assure pertinent QoS levels of the vital performance metrics, including the response time, a capacity planning methodology, based on the usage of relevant predictive models, has to be applied continually, as a crucial part of the e-Commerce Web site deployment. The solely aim of the process of capacity planning is to provide an unambiguous answer to the following question: 'Is the existing hardware infrastructure of a particular e-Commerce Web site capable to assure and maintain relevant Quality of Service (QoS) levels continually, having on mind the unpredictable and stochastic nature of e-Customer's online behavior?' In that context, Menascé & Almeida [4] define capacity planning as being '… the process of predicting when the future load levels will saturate the system and determining the most cost-effective way of delaying system saturation as much as possible', taking into account the natural evolution of the existing workload, the deployment of new applications and services, as well as the unpredictable and stochastic changes in e-Customer's behavior. Capacity planning requires appliance of predictive models, in order to make a prediction of the system's performance parameters, including: response times, throughput, network utilization, or resource queue lengths. All of these measures have to be estimated for a given set of known input parameters, including system parameters, resource parameters, and workload parameters. Contrary to trivial approach of applying non-regular, intuitive, *ad hoc* procedures that rely on arbitrary rules of thumb and personal experience, which usually lack proactive and continuous, scientifically based capacity planning methodology, a very systematic and thorough approach to addressing capacity planning issues, both on a system and component level, was developed by Menascé & Almeida [4] [5]. Their approach relies on utilization of the probability theory, the construction of a state transition graph, known as Customer Behavior Model Graph (CBMG) for a particular e-Commerce Web site, and the appliance of the theory of queues and queuing networks. The resulting predictive performance models, which consist of closed-form expressions for evaluating various performance metrics, offer analytical/numerical solution, and have been developed both on a system and component level. Yet another novel approach to capacity planning, based on modeling e-Customer's online behavior during e-Commerce shopping session, using the class of Deterministic and Stochastic Petri Nets (DSPNs), has been proposed by Mitrevski *et al.* [6]. It is suitable for obtaining various client-side related performance metrics, since it is also based on a CBMG of a particular e-Commerce Web site, and includes a stochastic temporal specification of the thinking times during e-Customer's stay in three main states: 'Browse', 'Search', and 'Checkout' (Fig. 1).

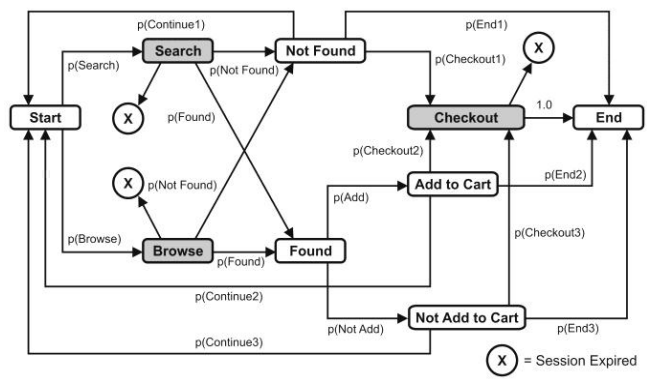

Fig. 1 CBMG corresponding to the DSPN model of e-Customer's online shopping behavior
(based on Mitrevski et al. [6])

The idea for utilizing the class of DSPN for modeling purposes is based on the fact that the underlying stochastic processes of both the e-Customer's online session and that one of the DSPN are, in fact, of the same type, i.e. Markov regenerative processes. This approach offers a possibility for obtaining both analytical and numerical solution, and the model can be also efficiently solved by specialized software. For the purposes of this paper, the DSPN model proposed by Mitrevski *et al.* [6], and represented via its CBMG (Fig. 1) will be utilized only as a template for implementing the logic of an e-Customer's online behavior within the discrete-event simulation model.

## II. DISCRETE-EVENT SIMULATION APPROACH TO CAPACITY PLANNING

Discrete-event simulation (DES) is one of the most widely used techniques for evaluating stochastic models. DES utilizes a mathematical/logical model of a physical system that portrays state changes at precise points in simulation time. Both the nature of the state changes and the time at which the changes occur, require precise description. Within DES, time advances not at equal size time steps, but rather until the next event can occur, so that the duration of activities determines how much the clock advances. Rather than employing any DES software simulator, we have used the open-source SimPy/Python programming environment for capacity planning purposes. SimPy (an acronym from 'Simulation in Python') is an extensible, object-oriented, process-based, general-purpose discrete-event simulation programming language, based on standard Python [7] [8] [9]. Compared to software simulators, and especially to all types of analytical/numerical solution methods, the usage of simulation programming language offers the greatest flexibility in terms of modeling power, and the maximum ability to depict an arbitrary level of details. For instance, all of the following aspects of the e-Commerce paradigm have been successfully modeled in SimPy: the client-side (the e-Customer's online behavior); the server-side (the hardware configuration of a typical e-Commerce Web site, on a system level); various classes of e-Customers (the qualitative component of the workload specification); the workload intensity (the quantitative component of the workload specification); the HTTP requests generation, for the three main functions invoked by the e-Customer: 'Browse', 'Search', and 'Checkout';         the propagation delays of the HTTP requests being forwarded from clients' browser towards e-Commerce servers and the delays of the corresponding responses, via Internet; and evaluation of plethora of performance metrics for both client- and server-side. Our SimPy implementation internally consists of three processes (named *Source*, *Customer*, and *Request*), along with their corresponding Process Execution Methods (PEMs) (Fig. 2).

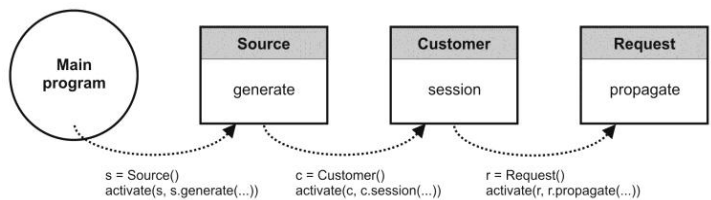

Fig. 2 A schematic representation of the internal structure of the SimPy simulation model

The *Source* process implements the generation of e-Customers according to the given arrival rate, and according to the probability distribution of their type. It also defines the parameters of the simulation runs. The *Customer* process implements the client-side, i.e. the e-Customer's online behavior during the online session, in accordance with the CBMG, depicted on Fig. 1. As stated previously, an e-Customer spends an arbitrary time, exponentially distributed, while being in three particular states, including 'Browse', 'Search', and 'Checkout'. The actual thinking times for those operations are being drawn from the same exponential distributions based on the average thinking times, i.e. 1.0, 1.0, and 3.0 minutes, respectively, for all classes of e-Customers, since we assume that e-Customers are all moderately experienced and need approximately the same amount of time whenever they perform those operations. If the e-Customers' segmentation into classes was based, for instance, on their experience (e.g. non-experienced, moderately experienced, and highly experienced), instead on their shopping behavior, the average thinking times for the 'Browse', 'Search' and 'Checkout' operations would be class-dependent, i.e. the non-experienced e-Customers would be assigned considerably bigger thinking times than the highly experienced ones. Whenever an e-Customer visits those states, a corresponding HTTP request has been generated and directed towards the e-Commerce Web site. Finally, the *Request* process implements the server-side, i.e. the propagation of each particular HTTP request through the e-Commerce Web site's hardware infrastructure, being already specified. Each Web site's server has been modeled on a system level, rather than on a component level. Since HTTP requests are processed by a particular server in a FIFO (FCFS) manner, each of them has been modeled as a resource with an infinite queue length, for simplicity reasons.

### III. WORKLOAD CHARACTERIZATION

The performance of any distributed system, like an e-Commerce Web site, which incorporates many clients, servers, and networks, depends heavily on the characteristics of its workload. According to Menascé & Almeida, the workload of a system can be defined as a set of all inputs that the system receives from its environment during any given period of time [4]. We focus on the workload characterization of the client-side. One has to be aware of two fundamental facts: first, e-Customers are not mutually equal, having on mind their online behavior; second, e-Customers access the Web site and invoke specific e-Commerce functions in an unpredictable and stochastic manner. The first fact is related to the qualitative aspects of the workload characterization. Many studies, including [10], have shown that it is possible to distinguish among various classes of e-Customers, having minded their online behavior during shopping sessions. Due to simplicity reasons, we model three basic classes of e-Customers regarding the intensity of buying online, i.e. *Rare Shoppers*, *Ordinary Shoppers*, and *Frequent Shoppers*. The specifics of their online shopping behavior can be defined through specification of the probabilities within the CBMG (Fig. 1), as shown in Table I. A mixture of various classes of e-Customers can be specified by defining a discrete random variable along with its probability mass function (pmf). If there are $k$ disjoint classes of e-Customers identified, e.g. $(t_1, t_2, \ldots, t_k)$, then each of them can be assigned a corresponding probability from the pmf vector $(p_1, p_2, \ldots, p_k)$, such that $\sum_{i=1}^{k} p_i = 1$, as a measure of its particular participation within the workload mixture. For our proposed classification: ($t_1 = $ *Rare Shopper*, $t_2 = $ *Ordinary Shopper*, $t_3 = $ *Frequent Shopper*), we are going to investigate the performance as a result of the service demand caused by three possible operating scenarios: $S_1(p_1 = 10\%; p_2 = 30\%; p_3 = 60\%)$, $S_2(p_1 = 33\%; p_2 = 34\%, p_3 = 33\%)$, and $S_3(p_1 = 50\%; p_2 = 30\%; p_3 = 20\%)$. It is also worthy to point out the flexibility of the model, i.e. the fact that by varying the values of the probabilities within the Table I, it is possible to model a wide range of different classes of e-Customers. For instance, for younger or novel e-Customers, the probability $p$(Browse) would be higher than the probability $p$(Search), due to their lack of experience. Similarly, the probability $p$(End2) of ending the online session without paying, after putting an item in the shopping basket, will be considerably higher with reluctant e-Customers.

TABLE I
DEFINING VARIOUS CLASSES OF E-CUSTOMERS USING DSPN MODEL (CBMG) PROBABILITIES

| CBMG (DSPN model) probabilities | Rare Shoppers | Ordinary Shoppers | Frequent Shoppers |
|---|---|---|---|
| $p$(Browse) | 0.50 | 0.50 | 0.50 |
| $p$(Search) | $1 - p$(Browse) | $1 - p$(Browse) | $1 - p$(Browse) |
| $p$(Found) | 0.10 | 0.50 | 0.90 |
| $p$(Not Found) | $1 - p$(Found) | $1 - p$(Found) | $1 - p$(Found) |
| $p$(Add) | 0.10 | 0.50 | 0.90 |
| $p$(Not Add) | $1 - p$(Add) | $1 - p$(Add) | $1 - p$(Add) |
| $p$(Continue1) | 0.10 | 0.33 | 0.50 |
| $p$(Checkout1) | 0.10 | 0.34 | 0.45 |
| $p$(End1) | 0.80 | 0.33 | 0.05 |
| $p$(Continue2) | 0.10 | 0.33 | 0.50 |
| $p$(Checkout2) | 0.10 | 0.34 | 0.45 |
| $p$(End2) | 0.80 | 0.33 | 0.05 |
| $p$(Continue3) | 0.10 | 0.33 | 0.50 |
| $p$(Checkout3) | 0.10 | 0.34 | 0.45 |
| $p$(End3) | 0.80 | 0.33 | 0.05 |

The second fact is related to the quantitative aspects of the workload characterization. The arrival process of e-Customers is a Poisson process [11], since: there is a zero probability of two arrivals at exactly the same instant of time; the number of arrivals in the future is independent of what have happened in the past; the number of arrivals in the future is independent and identically distributed (i.i.d.) random variable over time, i.e. the process is stationary.

The inter-arrival times of a Poisson process comprise an i.i.d. random variable with an exponential distribution. Therefore, since the memoryless property of the exponential distribution holds at any instant of time, the expected time until the next arrival is a constant and is given by $1/\lambda$, where $\lambda$ is the arrival rate [e-Customers/s] [12]. If $\lambda$ is the overall arrival rate for a given operating scenario ($S_1$, $S_2$, or $S_3$), each of them comprising of a mixture of e-Customer's types ($t_1$, $t_2$, …, $t_k$), along with a corresponding probability mass function ($p_1$, $p_2$, …, $p_k$), then the arrival rate of each e-Customer's type is given by the product $\lambda \cdot p_i$ ($i = 1, 2, …, k$) [12].

## IV. MODELING THE SERVER-SIDE

E-Commerce sites, especially those of the leading Internet retailers, have multi-tier hardware architecture, consisting of multiple servers, distributed throughout two or more LAN segments. In such a way, the site's architecture becomes more scalable, more flexible, more reliable and highly available [13]. We have modelled the e-Commerce server-side on a system level, not on a component level, as a set of resources, each having a queue with an infinite capacity. We keep in mind the simplest possible 2-tier hardware architecture of a medium-to-large scale e-Commerce system, consisting of a Front-End Server (FES), a Web Server (WS), a Database Server (DbS), an Application Server (ApS) and an Authentication Server (AuS), distributed into two high-speed LAN segments (Fig. 3). Each HTTP request may need several operations, i.e. types of processing by some of the back-end servers, before completing. Moreover, a request may have to be processed more than once in a particular back-end server. Fig. 4 depicts the typical Client-Server Interaction Diagram (CSID) for the 'Search' HTTP request [14]. Table II shows which sequence of servers is being activated for each request type [15].

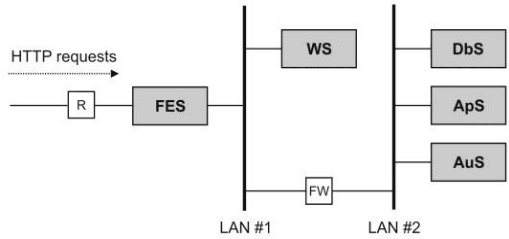

Fig. 3 Schematic representation of e-Commerce server-side hardware architecture

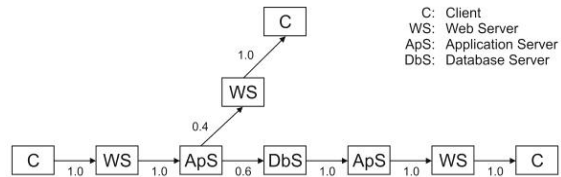

Fig. 4 CSID for the 'Search' e-Commerce function (Source: Menascé & Almeida [14])

TABLE II
A SEQUENCE OF BACK-END SERVERS ADDRESSED, FOR EACH SPECIFIC HTTP REQUEST'S TYPE

| Type of request | Back-end servers addressed |
|---|---|
| Search | WS, ApS, DbS, ApS, WS |
| Browse | WS, DbS, WS |
| Checkout | WS, AuS, DbS, AuS, WS |

Table III summarizes the different back-end servers [15], along with the mean processing time [ms] and supposed standard deviation ($\sigma$). We assume that the processing time has a Normal distribution and can vary up to $\pm10\%$ ($\pm3\sigma$) of its mean. In addition, the propagation time (transmission delay) via Internet WANs has been modeled as a random variable with a Normal distribution and parameters $N(\mu = 0.5$ [s]; $\sigma = 0.133333$ [s]), which yields 99.73% of the values within the interval $[0.1, \ldots, 0.9]$ [s]. At the server-side, the transmission delays between the servers have been neglected, due to the usage of high-speed LAN segments, as well as the propagation times in routers and firewalls.

TABLE III
SERVERS' STOCHASTIC PROCESSING TIME PARAMETERS

| Server | Mean processing time [ms] | Standard deviation [ms] | Range of values [ms] |
|---|---|---|---|
| Front-End Server (FES) | 1.0 | 3.33333E-05 | $[0.9, \ldots, 1.1]$ |
| Web Server (WS) | 10.0 | 3.33333E-04 | $[9.0, \ldots, 11.0]$ |
| Database Server (DbS) | 5.0 | 1.66666E-04 | $[4.5, \ldots, 5.5]$ |
| Application Server (ApS) | 10.0 | 3.33333E-04 | $[9.0, \ldots, 11.0]$ |
| Authentication Server (AuS) | 10.0 | 3.33333E-04 | $[9.0, \ldots, 11.0]$ |

## V. SIMULATION RESULTS

For each of the specified scenarios, a series of simulation runs have been carried out in order to estimate the values of various performance metrics, as a function of the arrival rates of e-Customers, ranging within the interval [0.0, …, 30.0] [s$^{-1}$] with a step of 0.5. Each run took into account a time window of 7200 seconds, or 2 hours of simulated time, long enough to get steady-state values of the estimated parameters. The response time is one of the most known and frequently used server-side performance metrics, which is of a vital interest to e-Customers. The average response time as a function of the e-Customer's arrival rate is given on Fig. 5a, for the three operating scenarios. As expected, Fig. 5a shows that the operating scenario $S_1$ generates a workload intensity that puts the biggest service demand on the Web site's infrastructure, since *Frequent Shoppers* are presented with 60%. Consequently, this means more repeated invocation of 'Add-to-Cart' and 'Checkout' functions, which pose greater number of HTTP requests to servers, especially to the Web server, queuing them to wait (Fig. 5b), thus resulting in bigger overall processing delays, decreased throughput (Fig. 5c) and decreased utilization (Fig. 5d). The estimated critical values of e-Customer's arrival rates for which the average response time reaches the 'psychological' threshold of 4.0 seconds for all three scenarios, have been obtained by using linear interpolation method, and they are, respectively, $\lambda_{S1}$ = 14.78; $\lambda_{S2}$ = 19.81; and $\lambda_{S3}$ = 24.28. The Web site's management strive is to retain at least those values, or, in the best case, to try to push them towards bigger ones, i.e. towards the right end of the axis, although during burst periods, e.g. holiday seasons, one can expect even higher percentage of *Frequent Shoppers* and consequently, a decrease of the critical values of arrival rates. To achieve this goal, relevant horizontal/vertical/diagonal scaling techniques of e-Commerce Web hardware infrastructure have to be applied on systems' component(s) that represent a bottleneck in the whole system, especially the Web server (WS). Regarding the experienced response time (RT), e-Customers have been categorized into three disjoint groups, including those experiencing a RT less than 2.0 seconds, e-Customers experiencing a RT between 2.0 and 4.0 seconds, and e-Customers experiencing a RT more than 4.0 seconds. The percentage of 'unhappy' e-Customers, i.e. those experiencing a RT lasting for more than 4.0 seconds, is increasing with the e-Customers' arrival rate, for all three operating scenarios. For instance, for a fixed $\lambda$ = 20 [e-Customers/s], the percentage of unhappy e-Customers is 98.69% (Scenario $S_1$), 67.66% (Scenario $S_2$) and 0.00% (Scenario $S_3$).

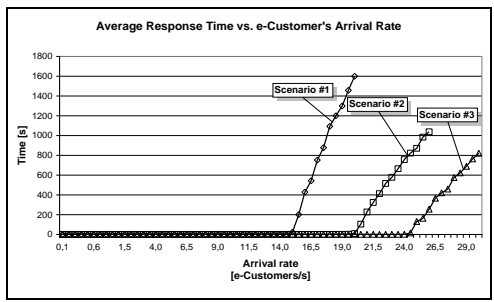

(a) Average response time vs. e-Customer's arrival rate

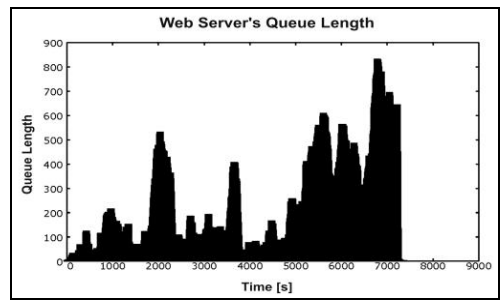

(b) The dynamics of Web server's queue length over time (Scenario $S_1$; $\lambda_{S1}$ = 14.78)

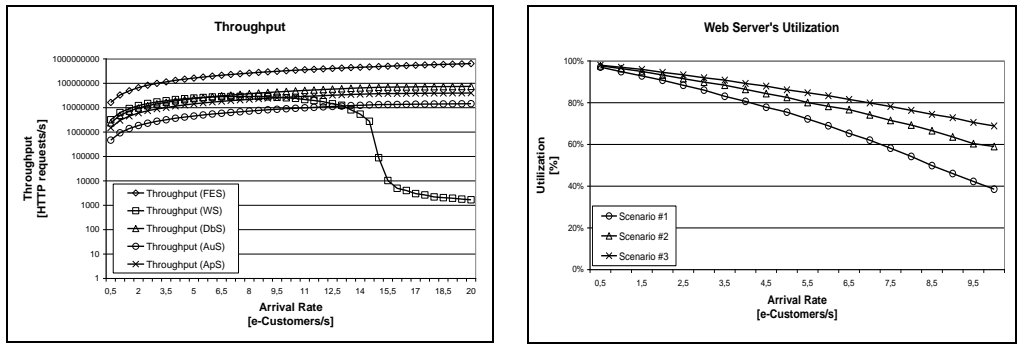

(c) Throughput of various server-side systems vs. e-Customers' arrival rate (Scenario $S_1$)

(d) Utilization [%] of the Web server (WS) for the three operating scenarios

Fig. 5 Simulation results

At client-side, besides numerous performance metrics that have been obtained, we first give those that can be also computed using the corresponding CBMG [4] [5] (Table IV):

- PM1: the average number of visits to each state, per session;
- PM2: steady-state probabilities of being in each state (except the 'Add-to-Cart' state);
- PM3: the percentage of customers that have left the site after having added at least one item into their shopping cart;
- PM4: the average session length [s];
- PM5: the 'Buy-to-Visit' ratio, showing how many sessions out of the total number have terminated with buying something.

We have also obtained the performance indicators, sublimed in Table V, including:

- PM6: the average sojourn time [s] spent in each state, per session;
- PM7: the percentage of sessions that have ended with an empty shopping cart;
- PM8: the average number of HTTP requests generated, per session;
- PM9: the percentage of e-Customers' sessions completed regularly;
- PM10/PM11: the average number of items being put in the shopping cart per session, both being paid for (PM10) and not being paid for (PM11). These performance measures can be used for estimating specific, business-oriented performance metrics, like *revenue throughput* [€/s] and *potential loss throughput* [€/s] [5].

TABLE IV
CLIENT-SIDE PERFORMANCE METRICS THAT CAN BE ALSO DERIVED DIRECTLY FROM THE CBMG

|  | Performance metrics | Scenario #1 | Scenario #2 | Scenario #3 |
|---|---|---|---|---|
| PM1 | 'Browse' | 0,84909 | 0,74141 | 0,68267 |
|  | 'Search' | 0,84773 | 0,74106 | 0,68317 |
|  | 'Add-to-Cart' | 1,03331 | 0,62983 | 0,41752 |
|  | 'Checkout' | 0,50874 | 0,30612 | 0,20076 |
| PM2 | 'Browse' | 0.26869 | 0.31117 | 0.35036 |
|  | 'Search' | 0.26897 | 0.31242 | 0.35155 |
|  | 'Checkout' | 0.46234 | 0.37641 | 0.29809 |
| PM3 | --- | $\cong 10.75\%$ | $\cong 8.93\%$ | $\cong 7.27\%$ |
| PM4 | --- | 189.4 [s] | 142.6 [s] | 116.3 [s] |
| PM5 | --- | $\cong 0.5$ | $\cong 0.3$ | $\cong 0.2$ |

TABLE V
ADDITIONAL CLIENT-SIDE PERFORMANCE METRICS

| Performance metrics | | Scenario #1 | Scenario #2 | Scenario #3 |
|---|---|---|---|---|
| PM6 | 'Browse' | 50.94720 | 44.54833 | 40.88380 |
| | 'Search' | 50.89374 | 44.37167 | 40.74863 |
| | 'Checkout' | 87.58523 | 53.67539 | 34.66748 |
| PM7 | --- | 19.62% | 40.62% | 53.82% |
| PM8 | --- | 3.20799 | 2.41300 | 1.976329 |
| PM9 | --- | 98.59% | 98.85% | 99.02% |
| PM10 | --- | 0.863179 | 0.507155 | 0.321171 |
| PM11 | --- | 0.157805 | 0.120595 | 0.093406 |

## VI. CONCLUSION

Discrete-event simulation by means of SimPy/Python code programming has proven to be an extremely powerful and flexible approach in modeling stochastic systems and/or processes like those already present with the contemporary e-Commerce paradigm. It allows one not only to assess a huge number of performance parameters that can be utilized for capacity planning, but also to experiment with various scenarios and hardware infrastructure's modalities, both on a system and component level. Still, the main obstacle with this approach is that it is both complex to implement and time-consuming. Regarding the substance in focus, the results obtained comply with those already given by Menascé & Almeida [4], whilst the DSPN model proposed by Mitrevski *et al.* [6] has proven to be a solid platform for capturing various e-Shoppers' behaviors. At server-side, future work includes performing a series of simulations in order to investigate the effects of a horizontal/vertical/diagonal scaling of the Web server, consideration of the availability and reliability (dependability) issues, and cost issues, as well. At client-side, future work includes verification of the performance metrics already obtained, both analytically/numerically, and by usage of dedicated software tools (TimeNET or DSPNExpress), and enrichment of the existing simulation model by including other general e-Commerce specific functions, in order to obtain more credible results. The simulation model can be also enhanced by introducing new classes of e-Customers, new performability metrics, as well as several other characteristic operating scenarios.